\documentclass[%
aip
 amsmath,amssymb,
reprint,%
superscriptaddress,
]{revtex4-2}

\usepackage{graphicx}
\usepackage{dcolumn}
\usepackage{bm}

\usepackage[utf8]{inputenc}
\usepackage[T1]{fontenc}
\usepackage{mathptmx}
\usepackage{etoolbox}

\begin{document}

\title{High-power laser beam shaping using a metasurface for shock excitation and focusing at the microscale}

\author{Yun Kai}
\email[Corresponding author: ]{ykai@mit.edu}
\affiliation{Department of Chemistry, Massachusetts Institute of Technology, Cambridge, MA 02139, USA}
\affiliation{Institute for Soldier Nanotechnologies, Massachusetts Institute of Technology, Cambridge, MA 02139, USA}
\author{Jet Lem}
\affiliation{Department of Chemistry, Massachusetts Institute of Technology, Cambridge, MA 02139, USA}
\affiliation{Institute for Soldier Nanotechnologies, Massachusetts Institute of Technology, Cambridge, MA 02139, USA}
\author{Marcus Ossiander}
\affiliation{John A. Paulson School of Engineering and Applied Sciences, Harvard University, Cambridge, MA 02138, USA}
\author{Maryna L. Meretska}
\affiliation{John A. Paulson School of Engineering and Applied Sciences, Harvard University, Cambridge, MA 02138, USA}
\author{Vyacheslav Sokurenko}
\affiliation{Kyiv Polytechnic Institute, National Technical University of Ukraine, 03056 Kyiv, Ukraine}
\author{Steven E. Kooi}
\affiliation{Institute for Soldier Nanotechnologies, Massachusetts Institute of Technology, Cambridge, MA 02139, USA}
\author{Federico Capasso}
\affiliation{John A. Paulson School of Engineering and Applied Sciences, Harvard University, Cambridge, MA 02138, USA}
\author{Keith A. Nelson}
\affiliation{Department of Chemistry, Massachusetts Institute of Technology, Cambridge, MA 02139, USA}
\affiliation{Institute for Soldier Nanotechnologies, Massachusetts Institute of Technology, Cambridge, MA 02139, USA}
\author{Thomas Pezeril}
\email[Corresponding author: ]{pezeril@mit.edu}
\affiliation{Department of Chemistry, Massachusetts Institute of Technology, Cambridge, MA 02139, USA}
\affiliation{Institut de Physique de Rennes, UMR CNRS 6251, Université Rennes 1, 35042 Rennes, France}

\begin{abstract}
Achieving high repeatability and efficiency in laser-induced strong shock wave excitation remains a significant technical challenge, as evidenced by the extensive efforts undertaken at large-scale national laboratories to optimize the compression of light element pellets. In this study, we propose and model a novel optical design for generating strong shocks at a tabletop scale. Our approach leverages the spatial and temporal shaping of multiple laser pulses to form concentric laser rings on condensed matter samples. Each laser ring initiates a two-dimensional focusing shock wave that overlaps and converges with preceding shock waves at a central point within the ring. We present preliminary experimental results for a single ring configuration. To enable high-power laser focusing at the micron scale, we demonstrate experimentally the feasibility of employing dielectric metasurfaces with exceptional damage threshold, experimentally determined to be 1.1 J/cm$^2$, as replacements for conventional optics. These metasurfaces enable the creation of pristine, high-fluence laser rings essential for launching stable shock waves in materials. Herein, we showcase results obtained using a water sample, achieving shock pressures in the gigapascal (GPa) range. Our findings provide a promising pathway towards the application of laser-induced strong shock compression in condensed matter at the microscale.
\end{abstract}

\date{\today}

\maketitle


Shock waves have broad significance across various fields, such as materials science \cite{veysset2015laser}, physical chemistry \cite{dlott2011new, li2022high}, astrophysics \cite{millot2019nanosecond}, medical therapies \cite{chaussy1980extracorporeally}, and more. In particular, information regarding high-pressure equations of state can be extracted through the study of shock propagation in condensed matter. In classical laser-shock experiments, shock waves are generated by delivering pulsed laser energy to a planar photoacoustic transducer layer deposited onto a sample. The absorption of laser energy induces an out-of-plane shock wave that travels into the sample. Commonly in these experiments, the shock wave is probed after propagation through the material at either a free surface or the interface of a transparent substrate. Such laser-induced shock wave techniques have typically been conducted at large-scale facilities, where high-energy laser pulses are readily available, often conveniently coupled with coherent x-ray sources or other techniques that enable advanced probing of the shock waves in samples. Given comparable shock durations of several nanoseconds in these experiments, the efficiency of laser-shock excitation can be quantified as the ratio of achieved shock pressure to the input laser pulse energy. Experimental campaigns conducted at the LCLS and OMEGA laser facilities have consistently revealed that the laser-shock excitation efficiency typically ranges from 15 GPa/J to 100 GPa/J \cite{mcbride2019phase,igumenshchev2017three}. Improving this efficiency is crucial for reaching the goal of achieving breakeven fusion.

\begin{figure*}[!hbt]
\centering\includegraphics[width=10cm]{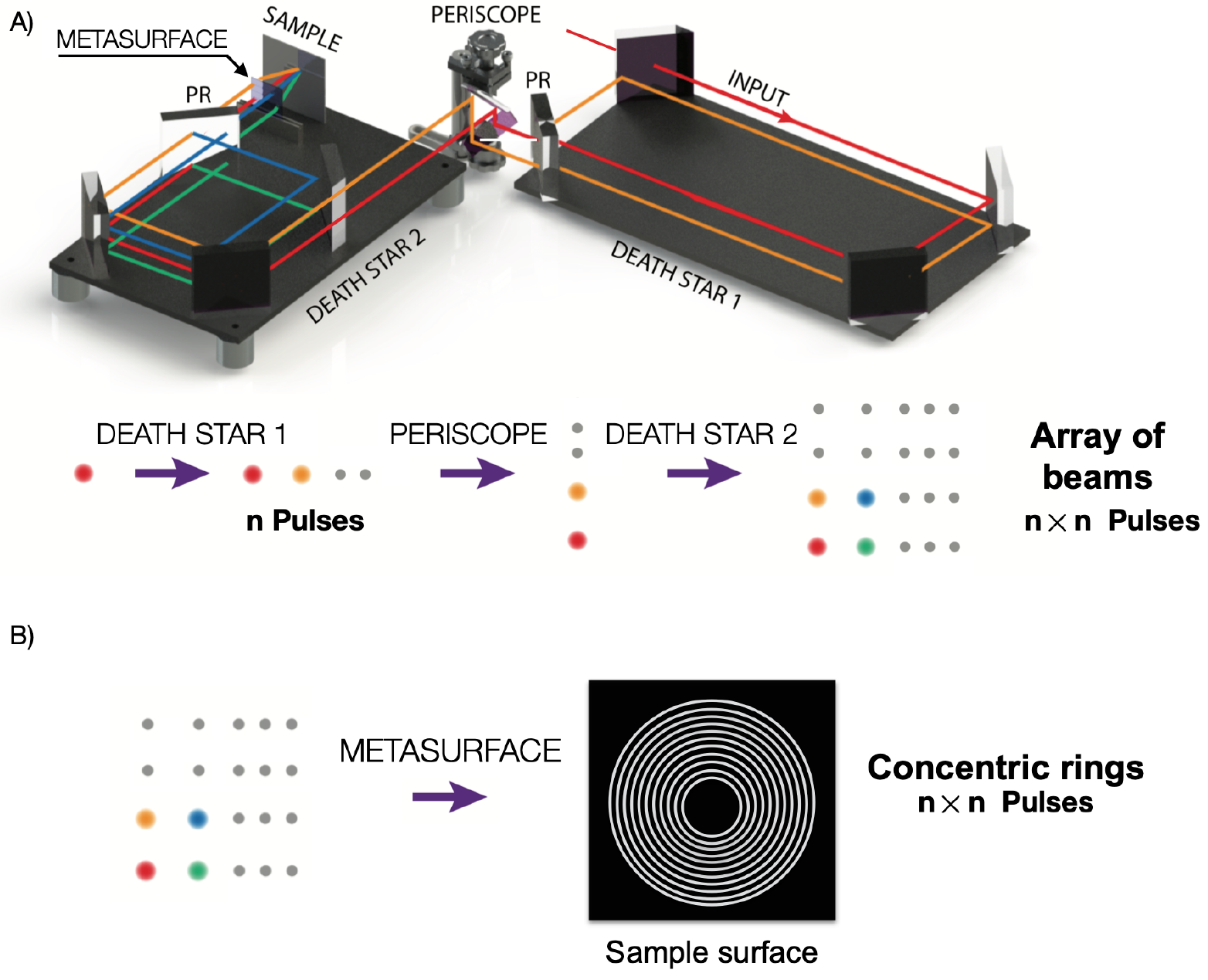}
\caption{(A) Dual Death Star setup to generate multiple time-delayed laser beams from a single input beam. The two “Death Star” cavities are formed of elementary square-shaped mirrors and a partial reflector (PR) for each. The sequential time delay between pulses is governed by the round-trip time of the Death Star cavity in the nanosecond range. (B) The matrix of n$\times$n collinear beams that emerges from the Dual Death Star setup can then be transmitted through a metasurface whose sophisticated 2D optical phase is calculated such that each of the beams focuses at the microscale on the sample surface and forms a concentric ring of a different diameter.}
\label{Fig_deathstar}
\end{figure*}

An alternative route towards shock wave experimentation is homogeneous direct-drive techniques. In these experiments, the sample itself serves as both the test material and shock launching layer by making the material of interest optically absorptive. Absorption of the laser energy leads to ablation of the material, launching shock waves diverging from the excitation region. This experimental geometry allows for the direct visualization of shock waves propagating laterally in the plane of the sample and is well-positioned for the application of spectroscopic probes to investigate shock-induced phenomena. Additionally, it allows the experimenter to shape the excitation arbitrarily for varied experimental designs. One such design demonstrated in our previous works\cite{pezeril2011direct, veysset2016interferometric, veysset2015laser}, is based on in-plane 2D focusing of shock waves. A converging shock wave is launched by shaping the laser pulse into a ring, whose pressure amplifies as it reaches the focus. This experimental geometry using a single laser ring has been shown to reach excitation efficiencies as high as $10^4$ GPa/J, with pressures of tens of GPa reached with pulse energies as low as a few mJs \cite{pezeril2011direct}. This allows one to conduct high-pressure experimentation with a commonplace low-cost tabletop laser amplifier system with the capability for rapid rearrangement and high throughput experimentation (hundreds of shots per day).

The herein-described novel optical design is expected to maintain or even exceed the shock excitation efficiency of $10^4$ GPa/J reported in our previous work. Indeed, thanks to much larger excitation areas with multiple laser ring excitations, we hope to bypass the saturation or plateauing effect encountered at increased laser fluences. The two novel excitation schemes that we describe herein, harness the fundamental principles of energy focusing and speed matching for optimized shock excitation efficiency. We have recently demonstrated the 1D proof-of-concept of spatial tuning and superposition of multiple weak shock waves excited by an array of laser photoacoustic sources shaped as lines ~\cite{Zebra}. In this pioneering experiments, we have shown that the optical device can excite $\times$20 higher shock pressures as compared to a single source that is limited by optical damage, and can maintain the high efficiency of the linear excitation regime. The basic idea of the novel optical design is based on the excitation of multiple synchronized laser rings rather than a single ring. The operating principle of this technique is the superposition of multiple converging shock waves. The first excitation will be the largest ring. The second excitation will excite a second ring with a smaller radius at a later inter-pulse time that matches the shock propagation time. In this way, the second ring shock is excited at the position of the previously launched first converging shock. This coherent multi-ring excitation is repeated for as many laser rings as available. Combining the shock velocity matching excitation and the 2D shock focusing should lead to a nonlinear build-up of the traveling shock pressure. Harnessing the power of both shock amplification through 2D focusing and shock velocity matching, one can expect significant increases in the achievable pressures \cite{patent}.

In this study, we present a novel optical design termed the dual “Death Star,” which enables the formation of an array of multiple time-delayed beams in the nanosecond range, matching the time scale of shock propagation in the sample. This array of temporally and spatially spread beams, generated by the dual “Death Star,” can be subsequently shaped and focused onto the sample as concentric rings at the microscale. To circumvent the use of refractive optics with low optical damage thresholds, which are unsuitable for high-energy pulsed lasers, we propose an alternative approach utilizing phase elements for shaping and focusing laser rings. Unlike conventional focusing elements, these phase elements alone could directly produce multiple rings at the sample location. While optical phase masks have been employed to focus or shape high-power laser beams at the millimeter scale \cite{dixit1994kinoform, ICF_review, Dlott}, their millimetric resolution is inadequate for microscale shock waves. In our work, we demonstrate the application of modern metasurfaces \cite{Metalens1, Metalens2, Marcus}, which exhibit a damage threshold of up to 1.1 J/cm$^2$, as a viable implementation for generating microscale laser-induced shock waves. Furthermore, we provide experimental results obtained from water samples, demonstrating the metasurfaces' ability to launch shock waves with pressures in the GPa range.

\section*{Multi-ring concept: Dual “Death Stars” optical design}

\begin{figure*}[!tb]
\centering\includegraphics[width=9cm]{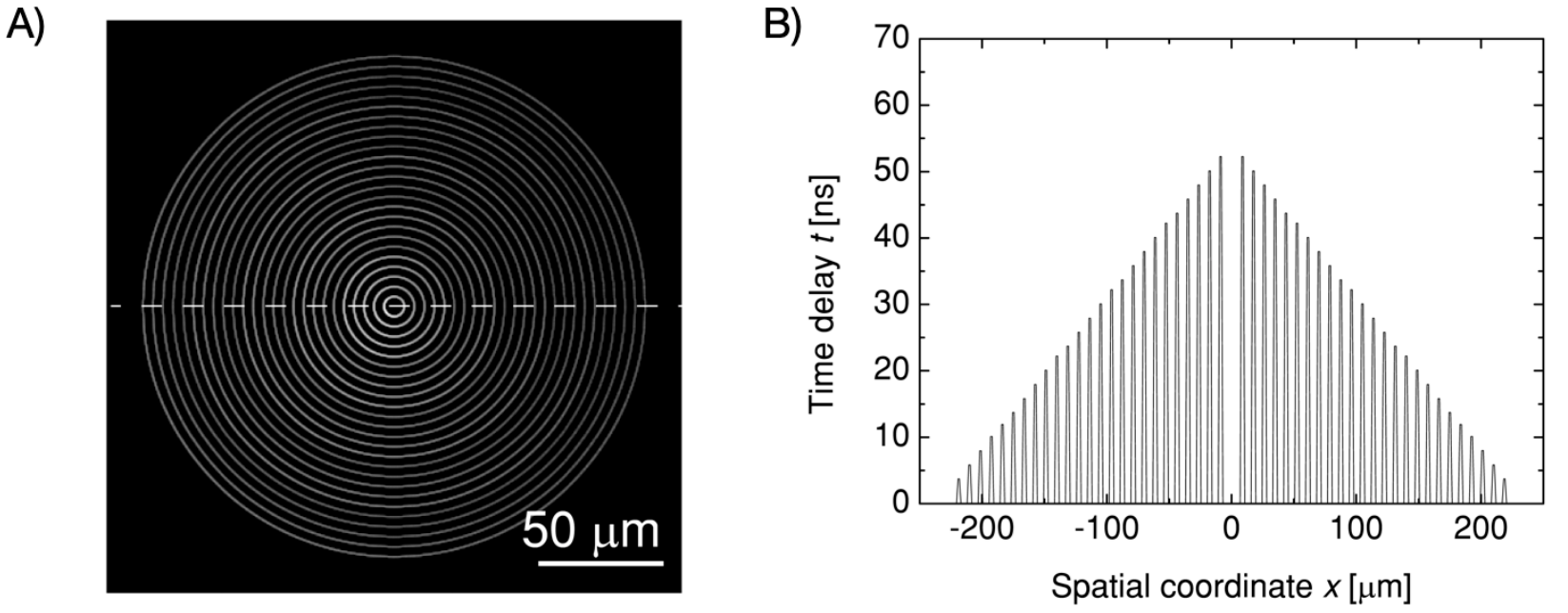}
\caption{(A) Simulated sample plane image at the focus of the metasurface, that incorporates the optical phase of a lens of 20~mm focal length, a matrix of axicon phases adapted to each of the input beams. The metasurface transforms the beam array into concentric rings, with each input beam generating a ring of different diameter. (B) Variation of time delay with spatial coordinate along the dashed line in (A), resulting in a laser scanning speed of 5~\textmu m/ns. Sequential timing of the rings ensures that the largest ring impacts the sample first, followed by subsequent rings of decreasing diameter until the smallest ring reaches the sample last.
}
\label{Fig_simu_deathstars}
\end{figure*}

The Death Star cavity, originally used for high-frequency (tens to hundreds of GHz) acoustic wave spectroscopy \cite{klieber2011narrow}, can be repurposed from its original use to obtain an array of beams with nanosecond inter-time delay or more. The Death Star operates with a four-mirror cyclic cavity, whose last mirror is a partial reflector (PR). After traveling through the cavity the PR allows part of the laser input pulse to exit the cavity and horizontally offsets the reflected laser beam for another round trip through the cavity. This leads to an output of n-horizontally spaced laser pulses, with temporal delay set by the Death Star round trip time. We will utilize two Death Star cavities connected by a twisted periscope in the proposed design. The first Death Star generates a horizontal array of n-pulses. The twisted periscope rotates this array by 90~degrees. This vertical array is input into the second Death Star cavity generating n-pulses from each input. In this way, we can generate a $n \times n$ matrix of pulses from an input single Gaussian pulse, see Fig.~\ref{Fig_deathstar}. Note that the laser intensity of each laser beam of the array depends on how the PRs composing the Death Stars are fragmented -- in practice, each PR is composed of many optical windows, laterally shifted such that each beam inside the cavity hits a specific window, with different optical transmission coefficients $T_n$. For instance, if each of the two partial reflectors PR is spatially fragmented with a change in transmittance that follows,
\begin{equation}
T_n= T_1, \,\, T_1/(1-T_1), \,\, T_1/(1-2T_1), ..., \,\, T_1/(1-(n-1)T_1), 
\label{Eq1}
\end{equation}
where $T_1$ is the optical transmittance for the first pulse arriving on the partial reflector, the intensity of each beam that exits the dual Death Star is identical. To limit losses on the last beam circulating the Death-Star cavity, the last transmission coefficient $T_n$ should be 100\%, which gives from Eq.(\ref{Eq1}) the extra condition on $T_1$, which is $T_1/(1-(n-1)T_1)=1$.

The laser beam array can then be transformed and focused into multiple rings at the sample surface by passing through a single element optical phase object \cite{lin2021coherent}. The Death Star setup's simulated laser beam profiles from a home-made ray-tracing optic software “Aber” developed at Kyiv Polytechnic Institute are presented in Fig.~\ref{Fig_simu_deathstars}. In the simulations, the input laser beam is 2.5~mm in diameter, the inter-time delay $\tau$ between pulses is set to 2~ns, the metasurface mimics the optical phase of an axicon array that consists of 5$\times$5 refractive axicons with angles $\theta_n$ that vary from $\theta_1$~=~0.05$^\circ$ to $\theta_{25}$~=~1.25$^\circ$ in 0.05$^\circ$ steps with an overall dimension of 50~mm~$\times$~50~mm. The metasurface incorporates as well the optical phase of a refractive lens of 20~mm focal length in order to focus all the beams on the sample surface with a common optical center. As shown in Fig.~\ref{Fig_simu_deathstars}(A) and (B), the conjunction of the optical phase of an axicon array and of an objective lens can generate multiple concentric rings with ring radii $r_n$~=~$220-10(n-1)$~\textmu m and constant ring width of 3~\textmu m, see Fig.~\ref{Fig_simu_deathstars}(A). Note that since each beam from the dual Death Star has equal intensity, the laser fluence of the rings in this example increases as 1/$r_n$ -- the ring surface is proportional to $r_n$. As mentioned above, it is possible to obtain any desired fluence distribution of the rings by adjusting the transmission coefficients $T_n$ in Eq.~(\ref{Eq1}). The laser scanning speed $v_n$, defined as the distance between adjacent rings $dr_n$ divided by the time delay $\tau$, in the simulation is 5~\textmu m/ns, see Fig.~\ref{Fig_simu_deathstars}(B), which is in range of the acoustic speed of many solid materials. A different set of focusing optics, axicon angles, and Death Star inter-time delay can be calculated to obtain a different laser scanning speed for varied samples. Owing to the nonlinear propagation of shock waves that imposes the shock speed to vary drastically with shock pressure, it is imperative to account for the nonlinear propagation of the shock waves in the sample. At very high laser energies and corresponding shock pressures, the laser scanning speed $v_n$ should not be constant anymore but should increase to follow the build-up of the shock wave. In this case, at the difference of Fig.~\ref{Fig_simu_deathstars}(A) calculated with a constant laser scanning speed $v_n$, the spatial separation of the rings $dr_n$ will not be a constant anymore but should rather shorten towards the center as the shock wave pressure builds up and increases in speed. To accommodate a changing shock speed, alternatively, we can adjust the time delay $\tau$ by without changing the spatial separation of the laser rings. This can be realized by adding delay lines at the output of each "Death Star" cavity.

\section*{Experimental results: Laser shock excitation in water using a metasurface}

\begin{figure*}[!tb]
\centering\includegraphics[width=9cm]{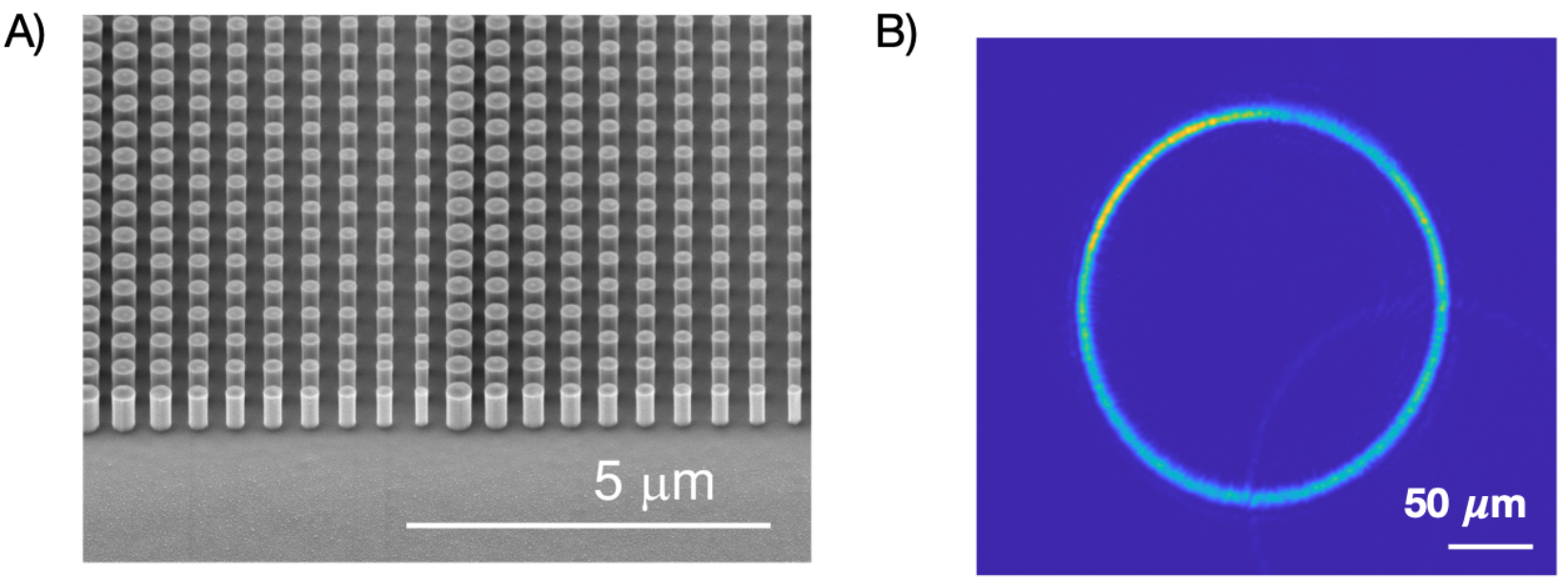}
\caption{(A) SEM image of the fabricated dielectric metasurface that incorporates the phase of a 0.5~$^\circ$ axicon phase and the phase of a lens of 20~mm focal length. (B) Recorded image of the beam profile at the sample plane, highlighting the laser ring generated by the metasurface. The result is a pristine 200~\textmu m laser ring, without any visible spurious hot spots, especially at the center.}
\label{Fig_metalens}
\end{figure*}

To demonstrate that modern metasurfaces \cite{Metalens1,Metalens2,Marcus} are a viable way for implementing this novel scheme for microscale laser-induced shock waves, we present laser-induced shock experiments utilizing a metasurface element to produce a single ring on the sample. The metasurface specifically fabricated for this study consists of 600~nm high circular titanium dioxide nanopillars arranged on a square 350~nm grid, see Fig.~\ref{Fig_metalens}(A). As this grid size is considerably smaller than the 532~nm operation wavelength, the metasurface only shapes the phase of the fundamental transmitted beam and creates no higher diffractive orders. By varying the nanopillar diameter, we emulate a digitized phase profile that combines the effect of a 0.5$^\circ$ axicon and a 10$\times$ objective lens of 20~mm focal length. The outcome reveals a pristine 200~\textmu m laser ring at the sample plane, devoid of any discernible spurious hot spots, particularly at the ring's central region where the presence of such unwanted light could significantly disrupt the shock focusing conditions, see Fig.~\ref{Fig_simu_deathstars}(B). Furthermore, to be able to induce strong shock waves, the metasurface must be able to sustain high laser fluences. Our investigations show no damage up to a fluence of $1.1~\text{J}/\text{cm}^2$ (tested by a 532~nm operation wavelength 5~ns pulsed YAG laser), higher than many dielectric mirrors.

We have conducted shock experiments following the methodology outlined in our seminal work \cite{pezeril2011direct}. In these experiments, a sub-ns pulsed laser (300~picoseconds duration, 532~nm wavelength, 5~Hz repetition rate, SL235 from Ekspla) was employed as the shock driver. To capture images of the sample at electronically tunable time delays, a femtosecond probe laser (200~fs duration, 400~nm wavelength, 1~kHz repetition rate, Legend from Coherent) was utilized. The sample configuration consisted of a thin water layer containing suspended carbon nanoparticles derived from ink (China Black Ink, Majuscule). The ink was diluted to achieve a nanoparticle loading in water of approximately 2 wt\%. The water layer was enclosed between two glass substrates of 100~\textmu m thickness each. A 25~\textmu m thick metallic ring spacer was inserted in order to control the water thickness. The shock wave was directly generated on a single shot within the sample by ultrafast absorption of the picosecond laser shock driver from the carbon nanoparticles. This transient absorption induced photoreactive energy release and vaporization of the nanoparticles, resulting in the generation of high-pressure conditions. The presence of the glass substrates created an impedance mismatch, effectively constraining the lateral expansion of the shock wave within the sample plane.

\begin{figure*}[!tb]
\centering\includegraphics[width=9cm]{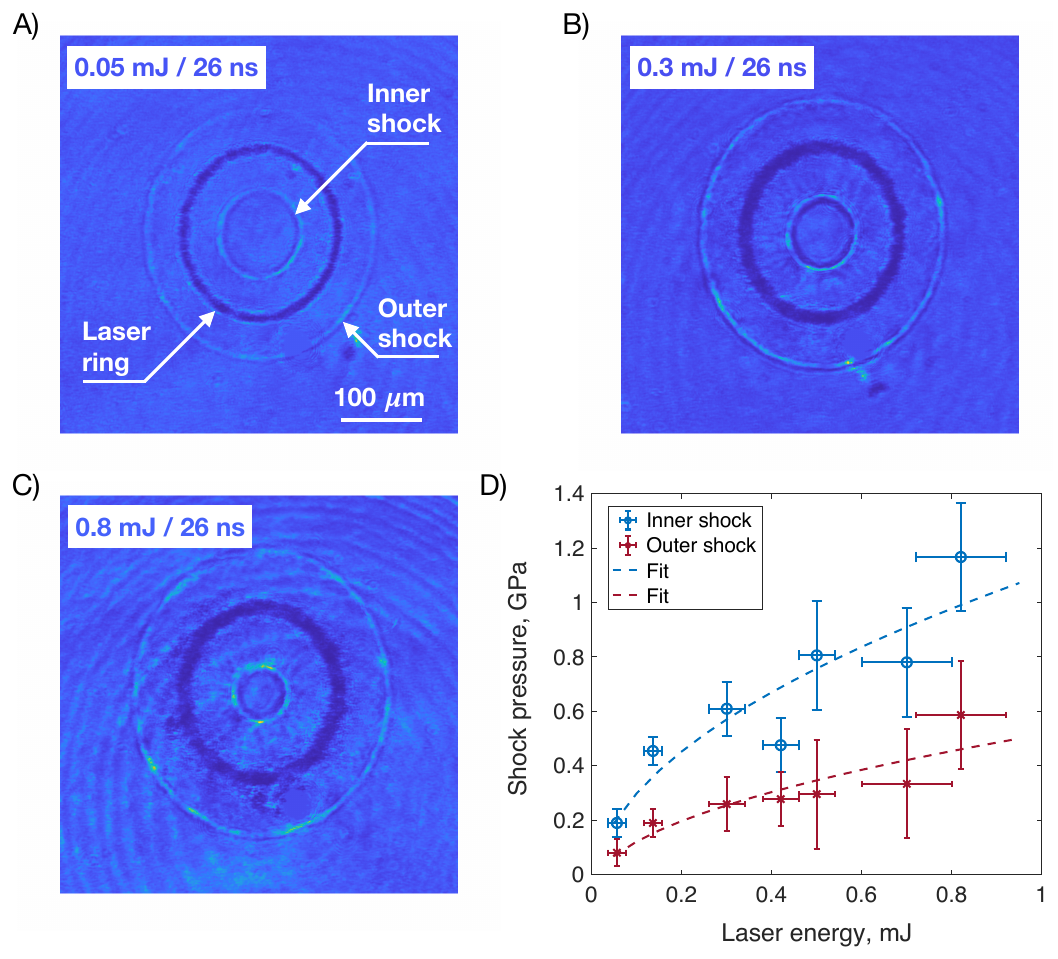}
\caption{(A)(B)(C) Single-shot recorded images at a ﬁxed time delay of 26~ns, taken at various optical shock pulse energies for a thin water sample. The inner shock launched by higher-energy pulses propagates farther toward the center of the ring due to higher shock pressures and, therefore higher speeds. (D) Plot of calculated shock pressure of the inner and outer shock waves as a function of laser pulse energy. The ﬁtted curve follows a square root function.}
\label{Fig_shocks}
\end{figure*}

The imaging setup employed allows for the recording of a single exposure at a specified probe pulse time delay each time the sample is irradiated, and a shock wave is generated. In Fig.~\ref{Fig_shocks}, images captured at a fixed time delay of 26~ns following irradiation with excitation pulses of varying energies are presented. These images, Fig.~\ref{Fig_shocks}(A) to (C), clearly highlight the increase in the propagation speed of the inner wave as the shock pulse energy is augmented from 0.05~mJ to 0.8~mJ. This behavior aligns with our expectations, as an increase in shock pressure typically corresponds to an increase in speed. Notably as well, the dark ring of bubbles surrounding the excitation region widens with higher laser pulse energy, serving as an indicator of the corresponding increase in pressure input. By analyzing these images, we can extract the average velocities of both the inner converging shock and the outer diverging shock. From our analysis, the inner shock speed $U_s$ increases from 1650~m/s to about 2400~m/s while the outer shock speed $U_s$ increases significantly less, from 1550~m/s to about 2000~m/s. Note that for the lowest laser pulse energy of 0.05~mJ, the outward propagating wave was close to the linear acoustic response limit corresponding in water to a speed equal to $c_0$~=~1.43~km/s. As in \cite{pezeril2011direct}, the shock-wave peak pressure $P$ in water can be calculated from the measured propagation speed $U_s$ of the shock wave through the equation of state at the shock front by,
\begin{equation}
P=\rho_0 \, U_s \frac{U_s - c_0}{1.99} \,\, \text{[GPa]}, 
\label{Eq1}
\end{equation}
where $\rho_0$~=~0.998~g/cm$^3$ denotes the density of the undisturbed water, and the factor of 1.99 was determined empirically. Thus, calculating the pressure from the speed is straightforward, yielding the results shown in Fig.~\ref{Fig_shocks}(D). At the highest laser pulse energy utilized, the inner shock wave exhibits a pressure of approximately 1~GPa, considerably higher than the outer shock wave, which has a pressure of roughly half that value. Note that this measured pressure is not the pressure at the shock focus that can be estimated from geometrical considerations to be about 20~GPa to 30~GPa. Given the quite high laser intensity fluctuation of the laser pumped by a flash lamp, the pressure versus laser energy curve in Fig.~\ref{Fig_shocks}(D) fit reasonably well with a square root function that captures the plateauing or saturation effect at increasing energy. These findings align perfectly with the results obtained in our previous work \cite{pezeril2011direct}, with the distinction that all refractive optics have been substituted with a metasurface. This advancement highlights the effectiveness of the metasurface in achieving similar shock wave dynamics at the microscale while eliminating the need for traditional refractive optical components not appropriate for high-energy lasers. 

\section*{Summary and outlook}
In our study, we have introduced an innovative optical cavity design that enables the generation of a concentric laser ring pattern with adjustable ring radii and time delay. We have conducted experimental demonstrations to showcase the crucial role of metasurfaces in shaping and focusing shock waves at the microscale. Building upon these advancements, we propose a potential follow-up approach wherein metasurfaces, which can now be manufactured on the wafer scale \cite{park2022all}, can be utilized to create a complete metasurface axicon and lens array. This implementation would transform the multi-beams generated by the novel optical cavity design into multi-rings, all in a single manufacturing process. We envision that this novel design will allow studying strong shock waves through the excitation and the merge of multiple 2D converging shock waves.

\section*{Acknowledgement}

We acknowledge funding from the DEVCOM Soldier Center and the Assistant Secretary of the Army for Acquisition Logistics and Training, specifically 0601102A, as well as the Laboratory University Collaboration Initiative (LUCI) program from the Department of Defense. T.P. acknowledges financial support from DGA (Direction Générale de l'Armement) under grant ERE 2017 60 0040, Rennes Métropole and Région Bretagne under grant SAD. Y.K. acknowledges support from a DAAD (German Academic Exchange Service) fellowship. This work was performed, in part, at the Center for Nanoscale Systems (CNS), a member of the National Nanotechnology Coordinated Infrastructure (NNCI), which is supported by the NSF under award no. ECCS-2025158. CNS is a part of Harvard University.

\bibliography{References}

\end{document}